# Towards a Usability Model for Software Development Process and Practice

Diego Fontdevila[1], Marcela Genero[2] and Alejandro Oliveros [1]

[1] Universidad Nacional de Tres de Febrero, Caseros, Argentina
{dfontdevila,aoliveros}@untref.edu.ar
[2] University of Castilla-La Mancha, Ciudad Real, Spain
marcela.genero@uclm.es

**Abstract.**
Context/Background: process and practice adoption is a key element in modern software process improvement initiatives, and many of them fail.
Goal: this paper presents a preliminary version of a usability model for software development process and practice.
Method: this model integrates different perspectives, the ISO Standard on Systems and Software Quality Models (ISO 25010) and classic usability literature. For illustrating the feasibility of the model, two experts applied it to Scrum.
Results: metrics values were mostly positive and consistent between evaluators.
Conclusions: we find the model feasible to use and potentially beneficial.

## 1 Introduction

Process and practice adoption is a key element in modern software process improvement initiatives, and it has become a central issue for organizations trying to become more agile. Many of these initiatives fail to accomplish their objectives [1][2], producing negative impact on costs, productivity and motivation for future improvements. On the other hand, there is evidence that human factors like emotion influence productivity, turnover, and job satisfaction in software development [3].

Processes and practices are tools that people use to coordinate and define their activities [4]; and adoption success may depend on the interactions between people as users of the process, and the process itself [5][6]. Since usability characterizes good interactions between users and tools that are appropriate and satisfactory to use [7], we propose that applying usability concepts to process and practice might improve adoption strategies. That is to say, focusing on process and practice usability might improve the probability of success of any process improvement, culture transformation or practice adoption initiative.

We initially defined process and practice usability as “how easy it is to follow a process or practice, including the effort needed to learn, the probability of making mistakes, the cost of such mistakes and the overall satisfaction and motivation pro-moted by following the practice or process." [8]. To operationalize this definition, our main contribution is the definition of a process and practice usability model composed of a set of sub-characteristics and metrics. Our model integrates three different sources, the work of Kroeger et al [5], the ISO Standard on Systems and Software Quality Models [7] and classic usability literature as well.

This model should help practitioners and process improvement specialists to better plan improvement initiatives, methodologists to better design new ways of working, and researchers to better understand adoption challenges. Adoption initiatives might increase their probability of success by adapting processes and practices to make them more usable, or at least by refining adoption strategies to take usability challenges into account. As an example, in the practice of Test Driven Development [9] the name of the practice suggests a testing practice but is actually about designing and coding software. Unclear naming is a typical usability issue [10].

The objective of this paper is to present a preliminary version of this model, and its application to Scrum as a feasibility study.

The rest of this paper is organized as follows: Section 2 presents related work, Section 3 describes our research method, Section 4 presents the preliminary version of the Usability Model for Software Development Process and Practice, Section 5 presents how we applied the model to Scrum, Section 6 analyzes threats to validity and Section 7 outlines the conclusions and future work.

## 2 Related Work

Very few studies consider people users of their processes or even mention process usability: Feiler and Humphrey mention process usability in the introduction to their work, but do not include it in their list of process quality attributes [11]. Culver-Lozo discusses usability but in terms of process documentation usability [12]. Kroeger et al have published significant research on the subject [5]. As an example of methodology analysis in terms of its relationship with its practitioners, Alistair Cockburn has reflected on the concept of high-discipline methodologies [13], which he defines as those that might probably be abandoned if a mechanism to keep them up is not put in place (an example of such mechanism is the Coach role in XP).

Kroeger et al [5] built their model from the concepts that they identified as quality attributes for software development processes. These quality attributes, in turn, the researchers grouped into 4 groups: Suitability, Usability, Manageability and Evolvability. They arrived at Usability as a grouping of: Learnability, Understandability, Accessibility and Adaptability. The ISO 25010 Standard on Systems and software quality models presents a product-oriented perspective on usability. Considering process to be like a software product is an analogy that other researchers have already used [11]. Since there is no Software Development Process Quality Standard, using the product standard seemed the right complement to the study by Kroeger et al [5].

The classic literature on usability represented by the work of Norman [10] and Nielsen [14] brought into the model very specific and rich terminology. An example of this is the generalization of the concept of appropriateness Recognizability from ISO 25010, aligned with the principle of affordance from Norman [10], into Self-evident Purpose.

## 3 Research Method

Our research includes the following activities: review the state of the art for software development process and practice usability; define a usability model for software development process and practice; perform a feasibility study to determine preliminary viability; refine the model and perform model validation.

To define the model we first identified the source literature related to process and practice usability. We conducted unstructured interviews with expert researchers on the subject[1] to identify candidate sources. From references provided by some of the experts we established three source types: process and practice usability, classic product usability literature, and product usability standards. We chose the study by Kroeger et al [5] as the reference source for process usability, and three reference sources on product usability [7][10][14]. Then, for each of the sources we added all elements to an initial candidate list of sub-characteristics. We proceeded to identify and group similar concepts, and then to purge the ones that did not seem to fit. We then refined or changed names in specific cases, mainly for clarification purposes. Finally, we added candidate metrics, some inspired from metrics defined in the sources, but mainly based in our experience with software process and practice adoption. The main author defined the model as described, and both other authors acted as reviewers of the model. We then performed a feasibility study on Scrum as described in Section 5. At this point we are planning further model refinement and validation (See more details in Appendix A, sections 1 to 4).

## 4 A Usability Model for Software Development Process and Practice

In this section we present an analysis of the sources and then describe the Process and Practice Usability Model.

### 4.1 Analysis of model sources

We based our model in the following sources: the study by Kroeger et al [5]; ISO 25010 [7], a standard for quality of systems and software products; and the classic works by Norman [10] and Nielsen [14].

Kroeger et al [5] have developed a model for improving software development processes from the perspective of the people involved. Their model is a generic quality model. Beyond its wider scope and its sound research methodology, their model has limitations regarding usability: although they define Process Usability as "ease with which a software engineering process can be *interpreted* and *performed* by practitioners" (the highlight is ours), its quality sub-attributes have little relationship with actual process performance. Specifically, understandability, learnability and accessibility are related to the interpretation of the process, and adaptability to is modification, which leaves no attribute to characterize process performance. Their definition of accessibility is "ease with which a process user is able to find information about a software engineering process" [5], which is focused in what we consider today a comparatively minor issue, information acquisition, as opposed to the traditional definition of "access for users with different capabilities" [7]. From our perspective, the most significant interactions are those between the people involved and the actual process, not between the people and the process definition documentation.

[1] Personal interviews with Eduardo Miranda, Laurie Williams and Mario Piatinni.

ISO 25010 [7] is a systems and software products quality standard, it has a comprehensive usability perspective that includes "soft" sub-characteristics like user satisfaction and user interface aesthetics. It defines usability as "degree to which a product or system can be used by specified users to achieve specified goals with effectiveness, efficiency and satisfaction in a specified context of use". It provided our work with a more modern perspective on usability (i.e. more related to user experience). It also defines three of the four sub-characteristics that Kroeger et al [5] consider for usability (learnability, adaptability and accessibility), although in the case of accessibility, with a very different meaning; and adaptability is considered a sub-characteristic of maintainability, not usability.

The classic usability literature [10][14] provided the first elements for the earliest forms of the model, starting with Feedback [10] and Tolerate mistakes [14]. It also provided some of the more nuanced sub-characteristics, like Affordance, which the ISO 25010 standard [7] confirmed with its own Appropriateness recognizability sub-characteristic. We later renamed affordance to Self-evident purpose, to increase model understandability since early discussions with expert practitioners[2] showed affordance as a term that was hard to apprehend.

### 4.2 The Model

The model is composed of nine sub-characteristics, which are aligned with our definition and emerged from the study of our model sources. In building the model we made sure that none of the concerns identified in the sources were left out, except accessibility as explained in Section 4.1, and avoid modes [14], which seemed inapplicable.

The model has several sub-characteristics that support process performance, in particular: visibility, that characterizes how transparent the status of a process and its intermediate products are to its stakeholders; controllability, that describes how easy it is for different stakeholders to control a process or practice during execution; and user satisfaction, which is a by-product of the experience of using the process or practice.

[2] Mary and Tom Poppendieck, Alistair Cockburn, Tobias Mayer and Brian Marick.

For each sub-characteristic we present a name, a definition and explain the rationale behind the inclusion of that sub-characteristic. We also present a set of candidate metrics for each sub-characteristic. The sub-characteristics are presented in Table 1 and the candidate metrics in Table 2.

**Table 1.** Process and practice usability sub-characteristics.

| Sub-characteristic | Definition | Rationale |
|---|---|---|
| Self-evident purpose | Degree to which users can recognize what a process or practice is for. | Purpose is a key motivator. Newcomers to a process or practice need to be able to make sense of it. |
| Learnability | "Ease with which a process user is able to learn how to perform the activities of a software engineering process." [5] | Difficulty to learn a new process or practice is a basic barrier for adoption. |
| Under-standability | "Ease with which a process user is able to understand whether a software engineering process is relevant and how it can be used to achieve desired results." [5] | Understandability applies to process and practice selection before adoption, and also during process performance. |
| Error tolerance | Degree to which the process is safe for its users, preventing errors or limiting their impact. | Error tolerance supports efficiency and effectiveness, and it also makes a process or practice easier to learn "on the job". |
| Visibility | Degree to which process structure, activities, status and information inputs and outputs are visible to stakeholders of the process in a specified context of use. | Visibility allows stakeholders to know the status of a process or practice and take early corrective action when necessary. It also helps set realistic expectations early. |
| Controllability | Degree to which a process or practice has attributes that make it easy to control. | Decisions need to be made at the appropriate time and impact the results effectively. |
| Adaptability | "ease with which a process user is able to adapt a software engineering process for use in different situations" [5] | Adaptability is about a process or practice supporting different contexts and users. This allows better fit and a higher reuse rate. |
| Attractiveness | Degree to which users of the process or practice find it attractive or resonate with its form or structure. | Attractiveness characterizes the appeal to newcomers. It might impact the desire to learn and adopt. |
| User satisfaction | Degree to which user needs are satisfied when using a process or practice | Satisfaction is a key element for positive feedback and impacts the creation of new habits |

To improve model application consistency and make it easier to use, we defined an evaluation process based on the ISO 25040 [15]. Table 2 describes model metrics.

**Table 2.** Candidate metrics.

| Sub-characteristic | Candidate Metric | Definition | Values | Type |
|---|---|---|---|---|
| Self-evident purpose | Appropriateness of name | Measures how appropriate the name is for describing the purpose of the process or practice. | Deceiving, Ambiguous, Partial, Appropriate, Accurate | Nominal |
| Self-evident purpose | Purpose alignment for stakeholders | Measures the alignment of purpose for all stakeholders. | None, Low, Medium, High, Complete | Ordinal |
| Learnability | Volume of information of introductory material | Measures the size of introductory material as defined by authoritative sources, e.g. for an authoritative introductory course. | Number of words | Absolute |
| Learnability | Standard introductory course duration | Measures standard course duration in hours, as defined by authoritative sources. | Number of hours | Absolute |
| Understandability | # of elements | Measures how many components make up the definition of the process or practice. | Number of elements | Absolute |
| Understandability | Conceptual model correspondence | Measures the level of correspondence between the user's conceptual model of an activity and the conceptual model of that same activity that the process or practice implies. | Low, Medium, High | Ordinal |
| Understandability | Data model complexity index | Measures the subjective complexity of the data model. | Low, Medium, High | Ordinal |
| Error tolerance | Cost of error | Measures the cost of error as overall impact. | Low, Medium, High | Ordinal |
| Error tolerance | Safety perception | Measures how safe is it to use the process or practice. | Low, Medium, High | Ordinal |
| Error tolerance | Use of restraining functions | Measures whether the process or practice provides hard restrictions to prevent risk materialization. | Yes/No | Nominal |
| Visibility | # of indicators | Measures how many standard indicators the process or practice defines. | Number of indicators | Absolute |
| Visibility | Use of information radiators | Measures whether information radiators are used in the process or practice. Information radiators display information regardless of user action. | Yes/No | Nominal |
| Visibility | Audience alignment for information | Measures whether information is presented in the same way to all stakeholders. | Yes/No | Nominal |
| Controllability | Degree of control concentration by role | Measures how concentrated control is among the roles defined. | Low, Medium, High | Ordinal |
| Controllability | Level of autonomy | Measures the level of autonomy users have in making decisions related to the process or practice. | Low, Medium, High | Ordinal |
| Controllability | Control granularity | Measures the control granularity of the process or practice. | Fine, Medium, Coarse | Ordinal |
| Adaptability | # of adaptation points | Measures how many adaptation points the process or practice defines. | Number of adaptation points | Absolute |
| Adaptability | Ratio of roles allowed to adapt | Measures how many roles are allowed to adapt the process or practice out of the total number of roles. | 0 to 1 | Ratio |
| Attractiveness | User atractiveness rating | Measures how attractive the process or practice is to prospective users (i.e. those lacking experience). | 1 to 5 | Ordinal |
| User satisfaction | User experience rating | Measures the subjective experience of using the process or practice. | 1 to 5 | Ordinal |

## 5 Applying the Model to Scrum

In this section we describe how we applied the model to Scrum to evaluate its feasibility. We limited evaluation to standard Scrum implementations [16]. First, one of the authors performed an evaluation, and then we proceeded to select two external Scrum experts[3] with more than 10 years of experience with Scrum. We provided them with introductory training to understand the model and the evaluation process, and also specific clarifications when required. For each model sub-characteristic, the evaluators assigned values to the model's candidate metrics, and added qualitative comments.

Evaluation results show that almost all metric values are in the middle or positive spectrum for that metric (see details in Appendix A, section 5). This is consistent with Scrum's popularity, simplicity and its focus on visibility and risk mitigation.

After the evaluation, informal feedback from the external evaluators provided interesting insights: granularity of the object of evaluation might be an issue (scrum vs. retrospective); differences between correct and incorrect implementations (one of the evaluators made a related distinction when evaluating Cost of error); distinguish standard from typical implementations (this emerged in the case of the Use of information radiators metric); evaluation is context sensitive (the Safety perception metric yielded two different values but with coherent underlying explanations); there are definitions that need to be improved. Overall, the results of both evaluators were highly consistent (see details in Appendix A, sections 5 and 6).

Finally, external evaluators were able to use the model effectively and produce qualitative comments that are aligned with model concepts. Thus, this provides initial confirmation that the model is understandable and feasible to apply.

## 6 Threats to Validity

Our work, being still on its early stages, presents issues that need to be addressed: lacks completeness validation, there is not enough confirmation of theoretical saturation; we cannot yet assess applicability to other processes or to specific practices; sample of evaluations is very limited, we have only two external evaluations; evaluators trained only with informal material (verbal explanations from the authors and access to the model in its current version); validation is limited, we need to improve on issues like consistency in evaluations by different evaluators and model accuracy in describing real life processes and practices.

## 7 Conclusions and Future Work

In this paper we presented our process and practice usability model, defining its sub-characteristics and candidate metrics. Through an initial application of the model to

[3] Juan Gabardini and Alan Cyment

Scrum by one author and two external evaluators, we found the model feasible to use and potentially beneficial.

Next steps include model and evaluation process refinement, including adding details, improving unclear definitions and metrics, defining how to compose metrics, and a user guide and training material; further validation with experts; application to other software development processes and practices to increase representativeness of the study; and empirical studies in industry.

**Appendix A:** Supplementary data available at https://doi.org/10.6084/m9.figshare.5296276.v1

## References

1. Ambler, S.: Agile practices survey results: July 2009, http://www.ambysoft.com/surveys/practices2009.html, last accessed 24-Jan-2017.
2. Paez, N., Fontdevila, D., Oliveros, A.: "Characterizing Technical and Organizational Practices in the Agile Community". In: Proc. CONAIISI, Salta, Argentina, (2016).
3. Graziotin, D., Wang, X., Abrahamsson, P.: "Software Developers, Moods, Emotions, and Performance", IEEE Software, July/August (2014).
4. Cockburn, A.: "What the Agile Toolbox Contains," Crosstalk Magazine, vol. Nov (2004).
5. Kroeger, T. A., Davidson, N. J., Cook, S. C.: "Understanding the characteristics of quality for software engineering processes: A Grounded Theory investigation", Information and Software Technology 56, pp 252–271, (2014)
6. Brown, J. S., Duguid, P.: The Social Life of Information. Harvard Business Press, (2000).
7. International Organization for Standardization, ISO/IEC 25010 Systems and Software Engineering - Systems and Software Quality Requirements and Evaluation (SQuaRE) - System and Software Quality Models. Geneva, Switzerland, (2011).
8. Fontdevila, D.: "A Tool Evaluation Framework based on fitness to Process and Practice," ICSEA, International Conference on Software Engineering Advances, Nice, France. (2014).
9. Beck, K.: Test Driven Development by Example. Boston, MA: Addison-Wesley, (2002).
10. Norman, D. A.: The design of everyday things. Basic books, (1988).
11. Feiler, P., Humphrey, W.: "Software process development and enactment: concepts and definitions", Software Engineering Institute, CMU/SEI-92-TR- 004, (1992).
12. Culver-Lozo, K.: "The software process from the developer's perspective: a case study on improving process usability," In: Proceedings. Ninth International Software Process Workshop, Airlie, VA, pp. 67-69. doi: 10.1109/ISPW.1994.512766 (1994).
13. Cockburn, A.: Agile Software Development: The Cooperative Game. Pearson Education, (2006).
14. Nielsen, J.: Usability Engineering. Elsevier, (1994).
15. International Organization for Standardization, ISO/IEC 25040 Systems and Software Engineering – System and software Quality Requirements and Evaluation (SQuaRE) – Evaluation process, Geneva, Switzerland, (2011).
16. Kchwaber, K., Sutherland, J.: Scrum Guide, http://www.scrumguides.org/scrum-guide.html, last accessed: 24-Jan-2017.